\documentclass[jpb,twocolumn,showpacs,superscriptaddress]{revtex4} 
\usepackage{graphicx}
\usepackage{amsmath}
\usepackage{amssymb}
\usepackage{epstopdf}
\usepackage{amsfonts}
\usepackage{dcolumn}
\usepackage{bigints}
\usepackage{xcolor}
\usepackage{hyperref}
\usepackage[normalem]{ulem}
\usepackage{csquotes}

\MakeOuterQuote{"}
\begin{document}
\title{Ro-vibrational Dynamics of the Neon Dimer}
\author{D. Blume}
\address{Homer L. Dodge Department of Physics and Astronomy,
  The University of Oklahoma,
  440 W. Brooks Street,
  Norman,
Oklahoma 73019, USA}
\address{Center for Quantum Research and Technology,
  The University of Oklahoma,
  440 W. Brooks Street,
  Norman,
Oklahoma 73019, USA}
\author{Q. Guan}
\address{Department of Physics and Astronomy, Washington State University, Pullman, Washington 99164-2814, USA}
\author{J. Kruse}
\address{Institut f\"ur Kernphysik, Goethe-Universit\"at; Frankfurt am Main, 60438, Germany}
\author{M. Kunitski}
\address{Institut f\"ur Kernphysik, Goethe-Universit\"at; Frankfurt am Main, 60438, Germany}
\author{R. D\"orner}
\address{Institut f\"ur Kernphysik, Goethe-Universit\"at; Frankfurt am Main, 60438, Germany}
\date{\today}

\begin{abstract}
Short intense laser pulses are routinely used to induce rotational wave packet dynamics of molecules. 
Ro-vibrational wave packet dynamics has been explored comparatively infrequently, focusing predominantly on extremely light 
and rigid molecules such as H$_2^+$, H$_2$, and D$_2$. This work presents quantum mechanical calculations that account for the rotational {\em{and}} the vibrational degrees of freedom for a heavier and rather floppy diatomic molecule, namely the neon dimer.  
For pumping by a strong and short non-resonant pump pulse, 
we identify several phenomena that depend critically on the vibrational (i.e., radial) degree of freedom. 
Our calculations show (i) 
fingerprints 
of the 
radial dynamics in the alignment signal; (ii) laser-kick induced dissociative dynamics on very short
time scales (ejection of highly structured "jets"); and
(iii)  tunneling dynamics that signifies the existence of resonance states, which are supported by the effective potential curves
for selected finite relative 
angular momenta.
Our theory predictions can be  
explored by 
existing state-of-the-art experiments. 
\end{abstract}
\maketitle

\section{Introduction}

Ultrafast spectroscopy, including pump-probe and pump-dump-probe 
spectroscopy, is an extremely powerful tool that has provided insights into
 electronic correlations of atoms and molecules embedded into liquids or solids 
as well as isolated atoms and molecules. 
While the spectral response provides enormous insights,
more recently direct imaging techniques have been developed for use, among others, in molecular
beam experiments~\cite{COLTRIMS,COLTRIMS2,schouder2022}. The ability to record
spatially and temporally resolved images experimentally with unprecedented 
precision opens the possibility to follow the pump-pulse induced dynamics 
of all degrees of freedom as a function of the delay time.

Pump-probe spectroscopy has, e.g., allowed for the creation of rotational wave packets that display unique revivals as a function of the delay time~\cite{RMP,RMP2,seideman1999,corkum2003,rotational,rotational2}. Such revivals have been observed in diatomic molecules as well as larger molecules for a wide range of pump pulse shapes and polarizations~\cite{lin2020}. In addition to being of fundamental relevance for understanding light-matter interactions, laser-kicked molecules promise a rich foray of applications, including sensing and high-harmonic generation~\cite{RMP}. 
Comparatively few studies of vibrational wave packets exist. Early work investigated the
spreading and recurrences of vibrational wave packets after electronic excitation or ionization~\cite{baumert}. Moreover,
 the distance-dependence of the ionization probability has been investigated in great
detail~\cite{distance-dependent-ionization,distance-dependent-ionization2}.
Vibrational or ro-vibrational wave packet dynamics have been studied  in light molecules such as H$_2^+$, D$_2^+$, H$_2$, and D$_2$,
where there exists an intriguing interplay also with the electronic degrees of 
freedom~\cite{rovibrational,rovibrational2,rovibrational3}.
An  
interplay of rotational and vibrational degrees of freedom is also seen in heavier
optically centrifuged super-rotors~\cite{villeneuve,centrifuge2}.  
Very recently, quantum control of the ro-vibrational dynamics has
been investigated in the context of light-induced molecular chirality in comparatively heavy molecules~\cite{koch2024}.

This work considers a simple linearly polarized Gaussian pump-pulse
with intensities and pulse length comparable to those used in many experiments
over the past several decades, namely
peak intensities $I$ up to $1.2 \times 10^{14}$~W/cm$^2$ and pulse lengths
up to $900$~fs.   
As an example, we consider the neon dimer. 
Compared to the helium dimer ground state, which is a weakly-bound quantum halo, the neon dimer ground state is
more strongly bound and more compact. Relatedly, the neon dimer possesses, unlike the helium dimer, several ro-vibrationally excited bound states. 
On the other hand, the characteristic rotational energy of the neon dimer is only about an order of magnitude smaller than
the vibrational energy scale, i.e., the scale separation is much smaller than in heavier rare gas dimers~\cite{hellmann_potential,rotational,heavydimer}.  
This comparatively small scale separation is an important prerequisite for identifying several novel phenomena that depend critically on the
internal dynamics of the molecule. 
A second ingredient decisive for the observed phenomena is the significant variation of the polarization anisotropy across the internuclear distances present in the neon dimer.
Specifically, we identify several phenomena
that 
are entirely absent in the rigid-rotor treatment~\cite{rigidbody}. 
We observe so-called Lochfrass, an effect that was first seen in seminal work on molecular hydrogen~\cite{saenz},
where the pump-laser triggers population transfer away from small internuclear distances, an effect that is absent in the rigid-rotor based description. 
Moreover, we observe unbound wave packet portions that fly away as "structured jets" with a speed of around a few
Bohr radii per picosecond. Wave packet portions that travel at much smaller speeds are a signature of  resonance states, which decay by quantum tunneling through the potential barrier created by the relative 
angular momentum of the dimer.
Furthermore, we highlight the distance-dependence of the alignment.

The remainder of this article is structured as follows. Section~\ref{sec_system} introduces the system under study and 
briefly introduces the theory framework employed. Section~\ref{sec_results} presents our results.
Finally, Sec.~\ref{sec_conclusion} provides a summary and outlook. Technical details are relegated to
Appendix~\ref{sec_appendix_rigid_rotor}.

\section{System Hamiltonian}
\label{sec_system}
The system Hamiltonian consists of the time-independent molecular part $H_0$ and the 
time-dependent laser-molecule
interaction $H_I(t)$, where $H_0$ accounts for the kinetic energy operator associated with the relative distance vector $\vec{R}$ and 
the isotropic neon-neon Born-Oppenheimer interaction potential $V_{\text{aa}}(R)$~\cite{hellmann_potential}; here $R=|\vec{R}|$. Throughout, we consider the dimer consisting of two
identical bosonic neon atoms ($^{20}$Ne) with  reduced
mass $\mu$, $\mu=18,221.99$~m$_e$ (m$_e$ denotes the electron mass), implying that the relative 
angular momentum
quantum number $J$ is restricted to even values by bosonic exchange symmetry. The dynamics is assumed to involve only the lowest electronic Born-Oppenheimer potential curve.

\begin{figure}
\vspace{-4.95in}
\hspace*{-0.2in}
\includegraphics[width=0.8\textwidth]{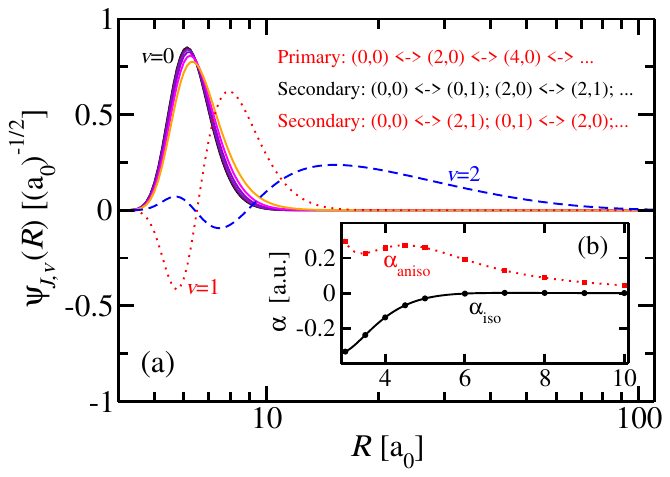}
\vspace*{-0.5in}
    \caption{(a) The solid lines show the scaled radial eigen functions $\psi_{J,0}(R)$ for $J=0$ (black solid line) to $J=10$ (orange solid line).
    For comparison, the red dotted and blue dashed lines show  $\psi_{0,v}(R)$ for $v=1$ and $2$, respectively.
    (b) Circles and squares show the  isotropic polarizability $\alpha_{\text{iso}}$ and anisotropic polarizability
    $\alpha_{\text{aniso}}$ from Refs.~\cite{polarizability1,polarizability2}. The lines show our fits.
    $J$-changing transitions require a non-zero  $\alpha_{\text{aniso}}$
    while $J$-preserving transitions require a non-zero  $\alpha_{\text{iso}}$ [see the color-coding of the text in the upper right corner of (a)].
    }
    \label{fig1}
\end{figure}

Table~\ref{tab1} shows that the
field-free neon dimer supports 6 bound vibrational ground states with energies $E_{J,v=0}$ ($J=0,2,\cdots,10$) and 4 
bound vibrationally excited states $\psi_{J,v}(R)$
with energies $E_{J,v>0}$. 
Solid, dotted, and dashed lines
in Fig.~\ref{fig1}(a) show the scaled radial eigen functions $\psi_{J,0}(R)$ for $v=0$ ($J=0,2,\cdots,10$), $v=1$ ($J=0$), and $v=2$ ($J=0$), respectively. The vibrational energy scale 
$E_{\text{vib}}$, $E_{\text{vib}}=E_{0,1}-E_{0,0}$, is about an order of magnitude larger than the rotational energy scale
$E_{\text{rot}}$, $E_{\text{rot}}=E_{2,0}-E_{0,0}$; this 
scale separation is less pronounced than that in heavier non-floppy molecules such as Ar$_2$~\cite{rotational} or I$_2$~\cite{iodine}.
Interestingly, the vibrationally excited states of the neon dimer are all quite close to the dissociation threshold, i.e., their binding energies are notably smaller than 
$E_{\text{vib}}$. 
Using the rotational constant $B$, the fifth column in Table~\ref{tab1} 
shows the energy $E_{J,\text{rigid}}$, $E_{J,\text{rigid}}=E_{0,0}+BJ(J+1)$, where 
$B$ is estimated using $\hbar^2/(2 \mu \langle R^2 \rangle)$, with the expectation value calculated with respect to $\psi_{0,0}(R)$. It can be seen that the rigid rotor model
provides a reasonable estimate of the rotational energies with $v=0$. Remarkably, it even provides a sensible estimate
of the energy of the resonance state that is supported by the $J=12$ channel~\cite{hellmann_potential}. 
For reference, Table~\ref{tab1} also reports the barrier height
$E_{J,\text{barrier}}$ of the
effective potential $V_{J,\text{eff}}(R)$, $V_{J,\text{eff}}(R)=V_{\text{aa}}(R)+\hbar^2 J(J+1)/(2 \mu R^2)$. 
The $R$-value at which $V_{J,\text{eff}}(R)$ takes on its local maximum changes from $R=18.69$~$a_0$ for $J=2$ to
$R=8.131$~$a_0$ for $J=14$ [the barrier height is measured relative to the value of $V_{J,\text{eff}}(R)$ for $R=\infty$].

\begin{table}
\begin{tabular}{cc|rrrr}
        $J$ & $v$  & $E_{J,v}$  [K] & $E_{J,v}$  [K]~\cite{hellmann_potential} & $E_{J,\text{rigid}}$ [K] & $E_{J,\text{barrier}}$ [K] \\ \hline
        $0$ & $0$ & $-24.0939$ & $-24.0941$ & $-24.0939$ & $0.0000$ \\ 
             $0$ & $1$ & $-4.2493$ & $-4.2494$ &  & \\ 
                  $0$ & $2$ & $-0.0188$ & $-0.0187$ &  & \\ \hline 
                       $2$ & $0$ & $-22.7504$ & $-22.7506$ &  $-22.7922$ & $0.0999$ \\ 
                            $2$ & $1$ & $-3.3838$ & $-3.3838$ &  & \\ \hline 
                                 $4$ & $0$ & $-19.6336$ & $-19.6338$ &  $-19.7549$ & $0.5977$ \\ 
                                      $4$ & $1$ & $-1.4399$ & $-1.4398$ &  & \\ \hline 
                                           $6$ & $0$ & $-14.7900$ &$-14.7901$ & $-14.9820$ & $1.7885$ \\ 
                                                $6$ & res. & & $1.2791$  & & $3.9473$\\ \hline 
                                                     $8$ & $0$ & $-8.3030$ & $-8.3031$ & $-8.4734$ & $3.9473$ \\  \hline 
                                                          $10$ & $0$ & $-0.3186$ & $-0.3187$ & $-0.2293$ & $7.3322$ \\  \hline 
                                                               $12$ & res. &  & $8.8362$ & $9.7505$ & $12.1977$ \\ \hline  
                                                                    $14$ & & unbound & $$ & $21.4658$  & $18.8179$ \\ 
        \end{tabular}
        \caption{Stationary bound states of the neon dimer. Our energies (third column) agree well with the literature (fourth column);
        "res."  stands for "resonance." 
        The rigid rotor energies are estimated using the expectation value $\langle R^2 \rangle$;
        the $J=0$ rigid rotor energy $E_{0,\text{rigid}}$ is equal to $E_{0,0}$ by construction.
     }
        \label{tab1}
        \end{table}

Letting the polarization vector lie along the laboratory-fixed
$z$-axis, the laser-molecule Hamiltonian reads~\cite{buckingham1973,potential_model,nielsen1999,polarizability1}  
\begin{eqnarray}
\label{eq_laser_molecule}
H_I(R,\theta,t)= \nonumber \\
-\frac{1}{2} |\epsilon(t)|^2
\left[ \alpha_{\text{iso}}(R) +  \frac{1}{3} \alpha_{\text{aniso}}(R)\sqrt{\frac{16 \pi}{5}} Y_{2,0}(\theta) \right],
\end{eqnarray}
where  $\theta$ denotes the angle  
that parametrizes 
the direction of $\vec{R}$ relative to the laser polarization.  
The spherical harmonic  $Y_{J,0}(\theta)$ is independent of the azimuthal angle $\phi$ 
since the projection quantum number
$M_J$ is equal to zero. 
Equation~(\ref{eq_laser_molecule}) includes an overall 
factor of $1/2$ that arises from averaging over the fast-oscillating electromagnetic field.
Solid and dotted lines in 
Fig.~\ref{fig1}(b) show
the $R$-dependent isotropic and anisotropic polarizabilities $\alpha_{\text{iso}}(R)$ and $\alpha_{\text{aniso}}(R)$, respectively~\cite{polarizability1}.  
The positive values of $\alpha_{\text{aniso}}(R)$ indicate that the energy is lowered when the molecular axis lies along the
$z$-axis [note the overall negative sign in Eq.~(\ref{eq_laser_molecule})]; this so-called head-to-tail arrangement is favored over the 
side-by-side arrangement ($\theta \approx \pi/2$) of the induced atomic dipoles. 
The laser-profile is parameterized through 
\begin{eqnarray}
|\epsilon(t)|^2 = |\overline{\epsilon}|^2 \exp \left[- 4 \text{ln} (2) \frac{(t-t_0)^2}{\tau^2} \right], 
\end{eqnarray}
where  $\tau$ denotes the pulse duration (FWHM) and
$t_0$ is a time offset [throughout, we use $t_0=\tau\sqrt{\ln(10^4)/\ln(2)}/2\approx 1.82 \tau$].
The peak intensity $I$ is given by 
$\epsilon_0 c |\overline{\epsilon}|^2/2$, 
where $c$ is the speed of light and $\epsilon_0$ the vacuum permittivity. 
Starting in an initial state with projection
quantum number $M_J=0$, the laser-molecule interaction populates only $M_J = 0$ states,
i.e., $M_J$ is a conserved quantum number (the system possesses an axial symmetry and the dynamics is independent of $\phi$).
Experimentally the induced dynamics can be observed by ionizing both atoms,
thereby  igniting a Coulomb explosion. This probe step is not modeled explicitly.

To solve the time-dependent Schr\"odinger equation, which depends---in addition to the time $t$---on $R$ and $\theta$,
we decompose the wave packet $\Psi(R,\theta,t)$ into  partial wave components
$u_J(R,t)$~\cite{helium_dimer1,helium_dimer2},
\begin{eqnarray}
\Psi(R,\theta,t) = \sum_{J=0,2,\cdots}^{J_{\text{max}}} \frac{u_J(R,t)}{R} Y_{J,0}(\theta).
\end{eqnarray}
The normalization
condition reads $\sum_{J=0,2,\cdots} P_J(t)=1$, where the channel populations $P_J(t)$ are given by $P_J(t)=\int_0^{\infty} |u_{J}(R,t)|^2 dR$.
We chose $u_0(R,0)=\psi_{0,0}(R)$ and $u_{J>0}(R,0)=0$, i.e., the initial state does not contain any $J>0$ contributions.
The time propagation is done by expanding the propagator into Chebychev polynomials~\cite{chebychev}. A linear grid in $R$ is used and convergence 
is ensured
by varying the time step, the number of polynomials considered in the propagator, the grid spacing in $R$, the box size
(minimum and maximum $R$ values), and the value of $J_{\text{max}}$.

\section{Results}
\label{sec_results}

\subsection{Alignment dynamics}
\begin{figure}
\vspace{-2.25in}
\hspace*{-0.in}
\includegraphics[width=0.6\textwidth]{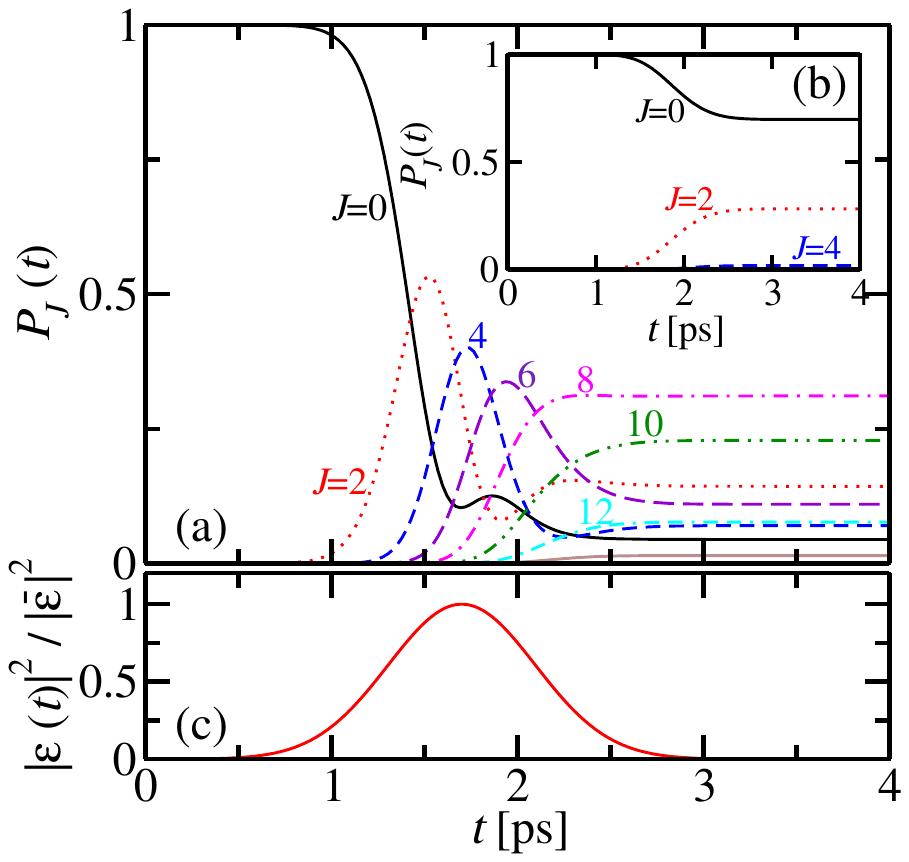}
\vspace*{-0.5in}
    \caption{
    Channel populations $P_J(t)$ as a function of time $t$ for a laser pulse with FWHM $\tau=933$~fs 
    and 
       (a)  $\overline{\epsilon}=0.0577$~a.u. (kick strength $P_{\text{kick}}=11.9$) and
     (b) $\overline{\epsilon}=0.0235$~a.u. ($P_{\text{kick}}=1.98$).
     $P_{\text{kick}}=11.9$ leads to the successive population of states with $J \lesssim 16$ [populations up to $J=12$
     are labeled in (a)] while  $P_{\text{kick}}=1.98$ leads to appreciable population
     of states with $J \lesssim 4$.
      The $P_J$ are constant after the laser pulse has decayed to essentially zero
      ($t \gtrsim 3$~ps).
      (c) The red solid line shows the normalized laser pulse for $\tau=933$~fs and $t_0 = 1.82 \tau$.
    }
    \label{fig2}
\end{figure}

\begin{figure}
\vspace{-3.2in}
\hspace*{-0.2in}
\includegraphics[width=0.8\textwidth]{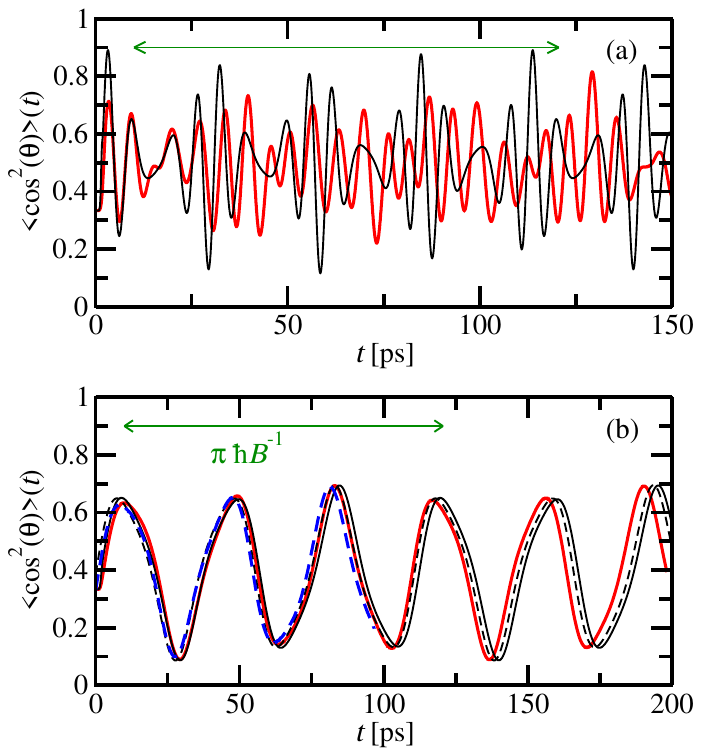}
\vspace*{-1.in}
    \caption{Alignment $ \langle \cos^2(\theta) \rangle (t)$ as a function of time for two different kick
    strengths $P_{\text{kick}}$, namely (a) $P_{\text{kick}}=11.9$ and (b) $P=1.98$. Thick (thin) lines show results where the radial dynamics is included
    (excluded).  
    In (b), the kick strength is realized using two different pulse lengths, namely $\tau = 933$~fs (solid lines) and 
    $\tau = 51.8$~fs (dashed lines).
    The green horizontal arrows in (a) and (b) indicate the  full revival period $T=\pi \hbar / B$ predicted by the rigid-rotor model.    
}
    \label{fig3}
\end{figure}

Figures~\ref{fig2}(a) and \ref{fig2}(b) show the channel populations $P_J(t)$ 
as a function of the delay time for  $\tau=933$~fs and  
a peak laser intensity $I$ of $I=1.17 \times 10^{14}$~W/cm$^2$ 
and $I=1.94 \times 10^{13}$~W/cm$^2$, respectively. 
For these parameter combinations it should be a reasonable assumption that the electronic degrees of freedom 
remain in the ground state during the dynamics~\cite{footnoteADK}.
The two parameter combinations correspond to dimensionless kick strengths
$P_{\text{kick}}=11.9$ and $1.98$, respectively, where~\cite{rigidbody,kickstrength2} 
\begin{eqnarray} 
P_{\text{kick}}=\frac{\langle \alpha_{\text{aniso}}(R) \rangle}{2 \hbar} \int_{-\infty}^{\infty} |\epsilon(t)|^2 dt. 
\end{eqnarray}
The quantity $P_{\text{kick}}/2$ provides a rough  estimate of the angular momentum transferred during the pump pulse.
Correspondingly, the two kick strengths lead to vastly different steady state channel 
populations after the pump pulse has decayed, at around $3$~ps, to a value close
to zero. The  pump pulse with $P_{\text{kick}}=1.98$ 
promotes nearly 30~\% of the population to the $J=2$ state and a negligible amount of population to channels with  $J>2$.
In contrast, the pump pulse with $P_{\text{kick}}=11.9$  sequentially populates channels with higher $J$ (first $J=2$, then $J=4$, etc.). 
This is a consequence of the fact that the transition matrix element from $(J,v)=(0,0)$ to $(2,0)$ is larger than the transition matrix elements
from $(0,0)$ to
$(0,v>0)$ and $(J>0,v>0)$; moreover, direct transitions from $(J,v)=(0,0)$ to states with $J>2$ are strictly forbidden [see also the upper right corner
of Fig.~\ref{fig1}(a)]. 
For the larger $P_{\text{kick}}$, about 25~\% of the population is, for $t \gtrsim 3$~ps, in the $J=8$ and $10$ channels each.
The $J=2$, $4$, $6$, and $12$ channels have around $10$~\% of the population each. 
We considered  various $\overline{\epsilon}$ and $\tau$ combinations (with $\tau$ as small as $50$~fs~\cite{footnote1}) 
and found 
that the dynamics for the same $P_{\text{kick}}$ are very similar,
regardless of the actual values of $\tau$ and $\overline{\epsilon}$ [this is illustrated explicitly in Fig.~\ref{fig3}(b) for one case]. This suggests that 
the laser pulse can, to a good approximation, be captured by a $\delta$-function kick~\cite{helium_dimer2}. The
$\delta$-function approximation does, of course, 
break down when the pulse length is comparable to the characteristic vibrational time scale of the neon dimer.

\begin{figure}
\vspace{-3.2in}
\hspace*{-0.2in}
\includegraphics[width=0.8\textwidth]{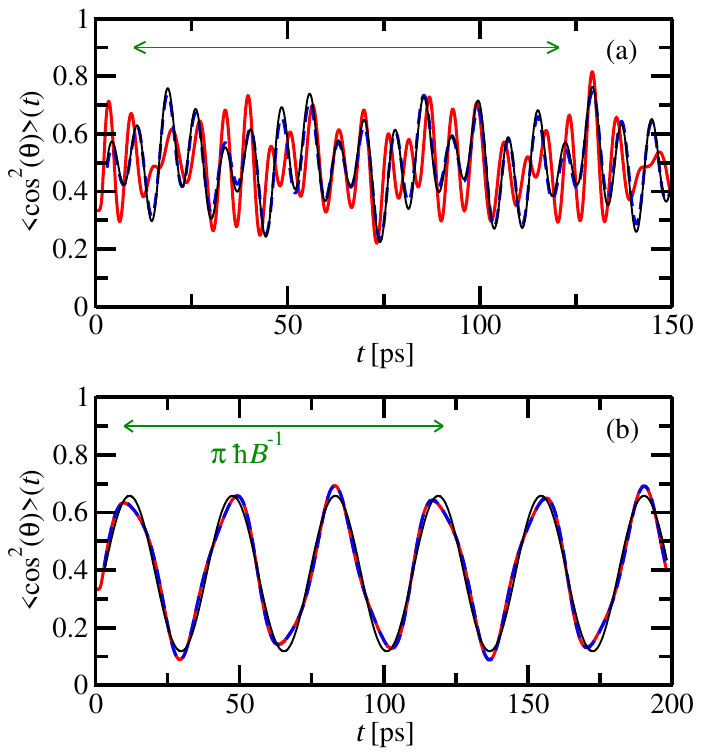}
\vspace*{-1.in}
    \caption{Constructing the alignment signal by projecting onto bound eigen states.
    The thick red curves are the same as shown in Fig.~\ref{fig3}.
    The thick blue dashed  and thin black solid lines are obtained by projecting the full wave packet at 3~ps onto the 
    bound eigen states and subsequently calculating the expectation value from the wave packet that is constructed using the eigen energies and eigen states.
    The calculations shown by the thick blue dashed line include all bound eigen states.
  The calculations shown by the thin black line, in contrast, only include  (a) the bound  eigen states  with $J=0,2,\cdots,10$ and $v=0$,
  and (b) the bound  eigen states  with  $J=0$ or $2$ and $v=0$.
    As in Fig.~\ref{fig3}, the green horizontal arrows in (a) and (b) indicate the  full revival period $T=\pi \hbar / B$ predicted by the rigid-rotor model.  
}
    \label{fig3_part2}
\end{figure}

\begin{figure}
\vspace{-.225in}
\hspace*{-0.45in}
\includegraphics[width=0.58\textwidth]{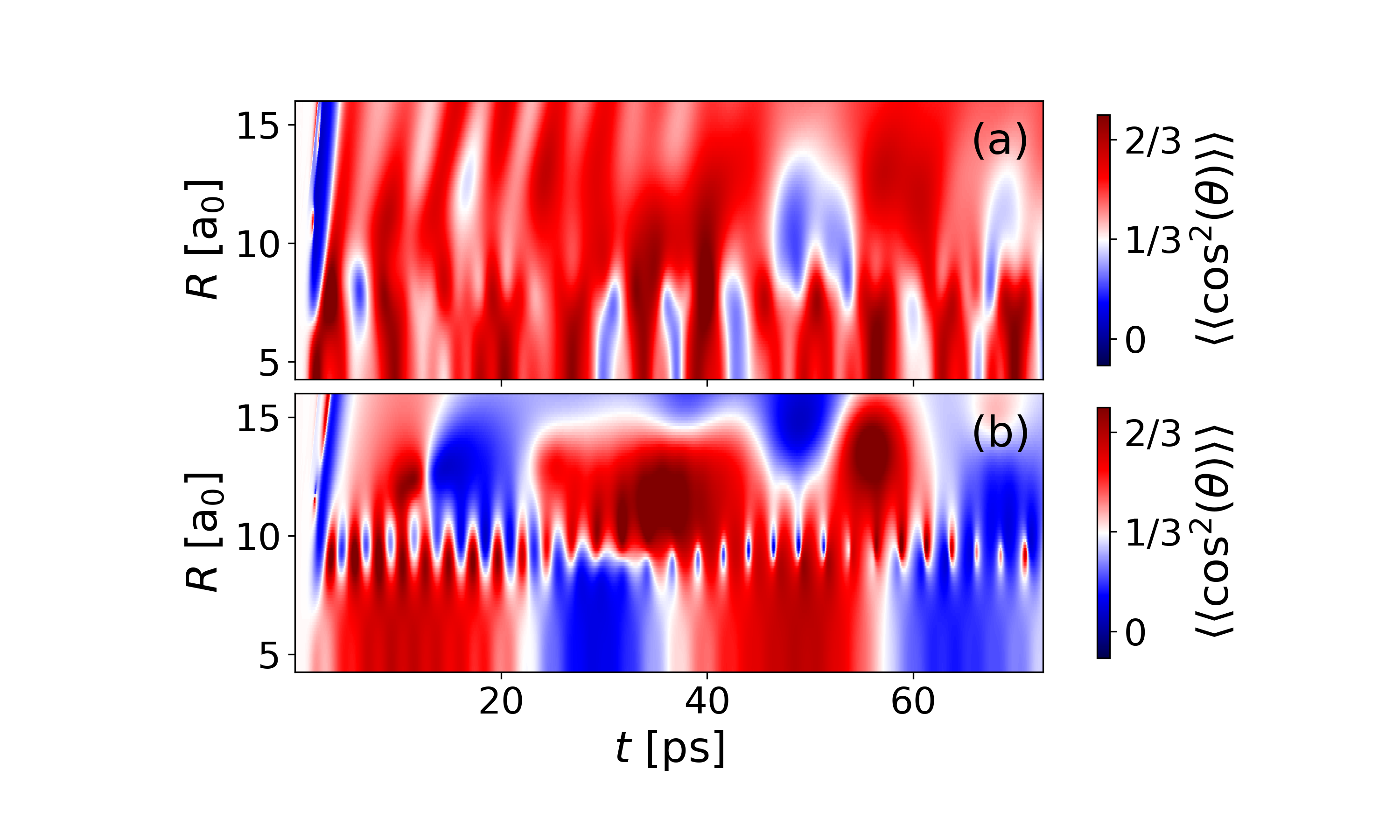}
\vspace*{-0.4in}
    \caption{Distance-resolved alignment $ \langle \langle \cos^2(\theta) \rangle \rangle (R,t)$ as a function of time for two different kick
    strengths $P_{\text{kick}}$, namely (a) $P_{\text{kick}}=11.9$  and (b) $P_{\text{kick}}=1.98$ ($\tau=933$~fs). 
    In looking at the figure, it should be kept in mind that the wave packet has a comparatively small amplitude for $R \gtrsim 8$~a$_0$; this can be deduced, e.g., by inspecting the eigen states (see Fig.~\ref{fig1}) and occupation probabilities (see Fig.~\ref{fig2}). 
    }
    \label{fig4}
\end{figure}

The supplemental material~\cite{SM} contains three movies that show the temporal evolution of the wave packet densities for $P_{\text{kick}}=11.9$ and $1.98$. Figures~\ref{fig3}-\ref{fig5} show observables that are extracted from the time-dependent wave packet. 
Red thick solid lines in Figs.~\ref{fig3}(a) and  \ref{fig3}(b) show the expectation value $\langle \cos^2(\theta) \rangle(t)$,
calculated with respect to the wave packet $\Psi(R,\theta,t)$, up to times that are significantly larger than $\tau=933$~fs for the same $P_{\text{kick}}$ as considered in Figs.~\ref{fig2}(a) and Figs.~\ref{fig2}(b). 
For both $P_{\text{kick}}$, the alignment signal increases at short times from its initial value of $1/3$ and then oscillates around a mean value [about $0.5$ for Fig.~\ref{fig3}(a) and about $0.4$ for Fig.~\ref{fig3}(b)], reaching values as 
large as $0.8$ and as small as $0.1$. 
Since the alignment signal is equal to $1/3$ for an isotropic density distribution, 
Fig.~\ref{fig3} shows an appreciable angular asymmetry of the density distribution for  $t \gg \tau$. Specifically, the 
angular asymmetry of the density distribution persists after the pulse has decayed to essentially zero. This phenomenon is known as  permanent alignment~\cite{RMP,RMP2}.
The physical picture is as follows: The laser pulse imparts energy into the neon dimer and creates a superposition of eigenstates, which is characterized by
a time averaged alignment signal that is larger than $1/3$.  This indicates that the superposition state created by the laser has an asymmetric density distribution, even if a time average is taken.    
Since the simulations are performed within a closed quantum system framework, the oscillations of the alignment signal are not damped and the  time-averaged alignment is larger than $1/3$ at all times. A more realistic modeling would include
decoherence due to coupling to the environment, which would reduce the oscillation amplitude with increasing delay time.
Typical experiments operate in regimes for which the decoherence time scale is larger than the delay times considered in Fig.~\ref{fig3}.

For the smaller $P_{\text{kick}}$ [Fig.~\ref{fig3}(b)], the oscillatory behavior is reminiscent of what one might expect for a two-state system 
without dissipation~\cite{rabimodel1,rabimodel2}. 
The quantity $\langle \cos^2 \theta \rangle(t)$ oscillates with a frequency that is, to a very good approximation, given by the energy difference
$E_{2,0}-E_{0,0} \approx 6B$.  
The thick blue dashed line shows the alignment signal for an 18 times shorter pulse ($\tau=51.8$~fs) but the same $P_{\text{kick}}$. The agreement between the two "full" calculations 
(calculations that account for the vibrational degrees of freedom) is excellent,
showcasing that the pulse length is, in this parameter regime, irrelevant for the dynamics. 
For comparison, the thin black
solid and  thin black dashed lines show results from the rigid rotor model~\cite{rigidbody,rigidbody3} (see Appendix~\ref{sec_appendix_rigid_rotor}). This model 
can be seen to capture the alignment dynamics convincingly.
Within the rigid rotor model, a complete revival occurs at $t+T$, where $T=\pi \hbar  /B$. Note that the period for the complete revival dynamics changes 
when even and odd $J$ are occupied.
For the neon dimer, 
$T$ is equal to $110.6$~ps. It can be seen that a full revival is observed (at least to a very good approximation) in this low kick strength regime even when the vibrational dynamics 
is accounted for explicitly.

For the larger $P_{\text{kick}}$ [Fig.~\ref{fig3}(a)], the alignment signal oscillates more rapidly and is governed by minimum and maximum amplitudes that
vary appreciably with time, indicating the presence of multiple time scales. 
While the rigid rotor model [thin solid line in Fig.~\ref{fig3}(a)] captures the overall behavior (namely, comparatively rapid oscillations), 
several differences can be observed. Specifically,
the rigid rotor model predicts larger amplitudes and fewer oscillations.
In the full description, the occupation of the bound $v>0$ eigen states is 4.5~\% for $t \gtrsim3$~ps. 
 Since the vibrational excitations $E_{J,1}-E_{J,0}$ are comparable to the rotational excitations 
$E_{J,0}-E_{J+J',0}$ with $J'$ around $8-10$, the larger $P_{\text{kick}}$ mixes the rotational and vibrational dynamics. 
The rigid rotor model, in contrast, has zero occupation in states with $v>0$. Because of this, the rigid rotor model predicts more regular oscillations of the alignment signal than the full calculation.
If the dynamics was fully captured by the rigid rotor
model, then full revivals should occur with the period $T$. The alignment calculated using the rigid rotor framework exhibits 
full revivals at period $T$ while the alignment calculated using the full quantum treatment does not.
This is very different than the dynamics of other rare gas dimers~\cite{mizuse2022,wu2011}.

Figure~\ref{fig3_part2} compares the alignment signal obtained from the full dynamics (thick red solid lines; the data are also shown in Fig.~\ref{fig3}) with that obtained by projecting the full wave packet onto the bound eigen states (thick blue dashed lines)
or a subset of the bound eigen states (thin black solid lines).
To obtain the thin black solid lines in Figs.~\ref{fig3_part2}(a)
and \ref{fig3_part2}(b), $v=0$ states with $J=0,2,\cdots, 10$ and $J=0, 2$, respectively, are included.
For $P_{\text{kick}}=1.98$ [Fig.~\ref{fig3_part2}(b)], the full wave packet dynamics is extremely well 
described by projecting onto the bound eigen states and even the inclusion of just two bound eigen states reproduces the alignment signal quite accurately.
 For $P_{\text{kick}}=11.9$ [Fig.~\ref{fig3_part2}(a)], the close agreement between the thick blue dashed
 and thin black solid lines indicates that the $v>0$ bound eigen states impact the dynamics comparatively weakly.
 There exists, however, an appreciable difference between the full wave packet dynamics
 and the dynamics captured by projecting onto the bound eigen states
 (compare the thick red solid and thick blue dashed lines). 
 The difference reflects that the full wave packet contains contributions from unbound
 resonance states (see Sec.~\ref{sec_resonance}).

To gain additional insight, we define the  $R$-resolved
alignment signal $\langle \langle \cos^2(\theta)\rangle \rangle(R,t)$,
where the double brackets indicate integration over
the $\theta$ degree of freedom only. At each $R$ value, the expectation value includes the normalization factor
$(\sum_{J=0,2,\cdots}|u_{J}(R,t)|^2)^{-1}$. For a spherically symmetric wave packet,  $\langle \langle \cos^2(\theta)\rangle \rangle(R,t)$
is equal to $1/3$ for all $R$, including $R$ values for which the density is low.
Figure~\ref{fig4} shows $\langle \langle \cos^2(\theta)\rangle \rangle(R,t)$ for two different
$P_{\text{kick}}$. Appreciable dynamics along the $R$ coordinate is observed.
Recall that the dynamics in $R$ is completely absent in the rigid-rotor framework.
The $R$-averaged alignment signal $\langle \cos^2(\theta)\rangle(t)$ (see Figs.~\ref{fig3} and \ref{fig3_part2}) is dominated by the   $R \lesssim 8$~$a_0$ region.
The "phase change" at $R \approx  8$~$a_0$ (this phase change can be seen particularly clearly in the 
lower peak strength data) is due to the node of the $v>0$ eigen states, which have low populations (see above). 
Even though the probability for $R>8$~$a_0$ is comparatively low, a dedicated experiment should be able to collect sufficient statistics to reveal the 
multi-scale dynamics that is encoded in the $R$-resolved alignment signal.

\subsection{Dynamics governed by resonance states}

\label{sec_resonance}

\begin{figure}
\vspace{-.5in}
\hspace*{-4in}
\includegraphics[width=1.2\textwidth]{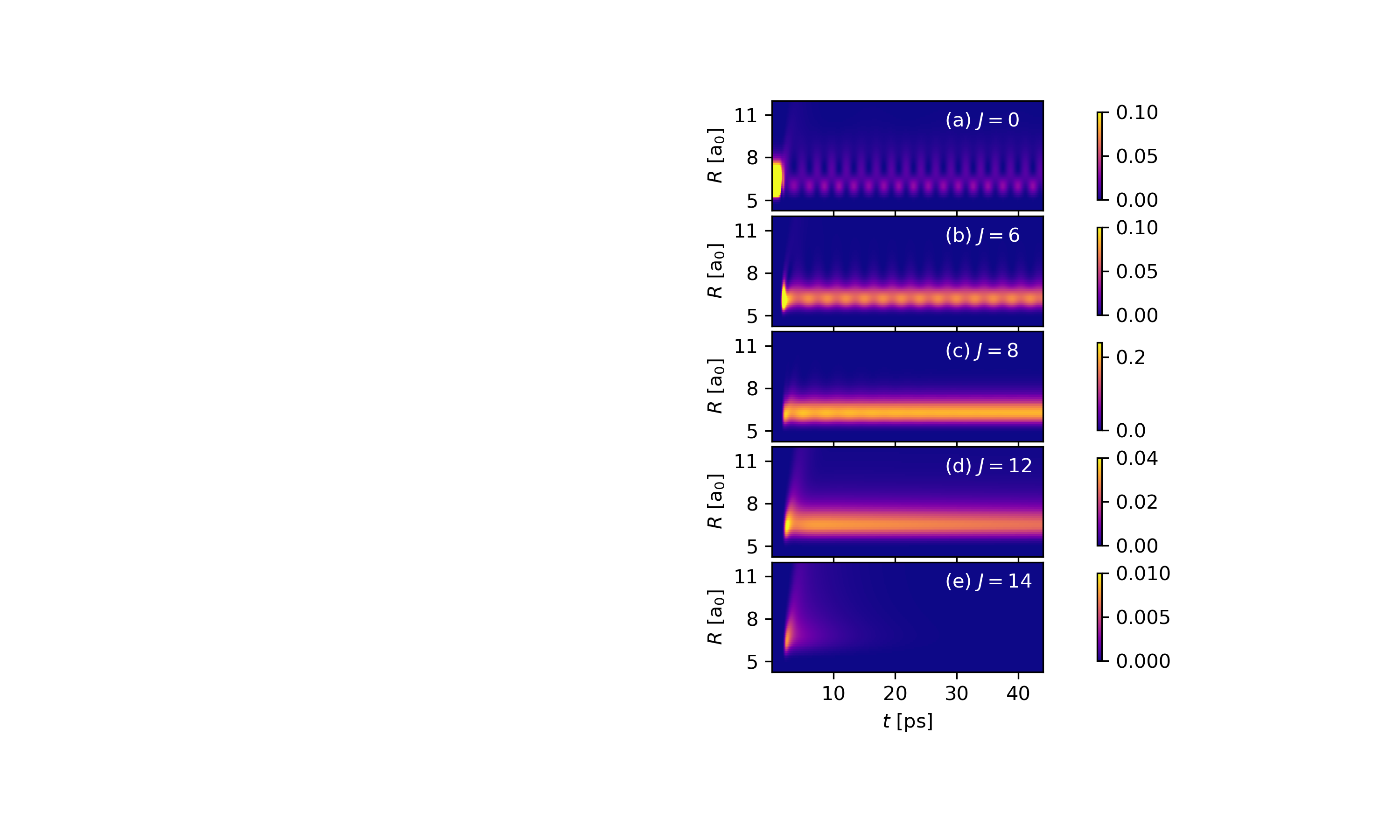}
\vspace*{-0.8in}
    \caption{Channel density $|u_{J}(R,t)|^2$ for 
    (a) $J=0$,
    (b) $J=6$,
    (c) $J=8$,
    (d) $J=12$, and
    (e) $J=14$
    as a function of time for a kick strength
    $P_{\text{kick}}=11.9$. 
    }
    \label{fig5}
\end{figure}

Figures~\ref{fig5}(a)-\ref{fig5}(e) show the channel densities $|u_{J}(R,t)|^2$ for $J=0$, $6$, $8$, $12$, and $14$
for $P_{\text{kick}}=11.9$ ($\tau=933$~fs). The laser pulse promotes a portion of the wave packet to
unbound scattering states, as is reflected in the fast wave packet portions ("jets") at small times that move at a speed of about a few Bohr radii per picosecond for all $J$ shown
(for $J=8$, the fast moving wave packet components are hard to see on the scale shown).
The occupation of unbound $J=0$ wavepacket portions is a manifestation of Lochfrass, i.e., "the non-uniform depletion 
of the wave packet" at small $R$~\cite{saenz}.
For $J=14$ [see Fig.~\ref{fig5}(e)], the entire channel density "flies away," i.e., the distance between the two neon atoms increases rapidly with time. We emphasize that the unbound wave packet portion is neutral 
(recall, ionization is excluded from our theoretical framework).

\begin{figure}
\vspace{-.2in}
\hspace*{.1in}
\includegraphics[width=0.45\textwidth]{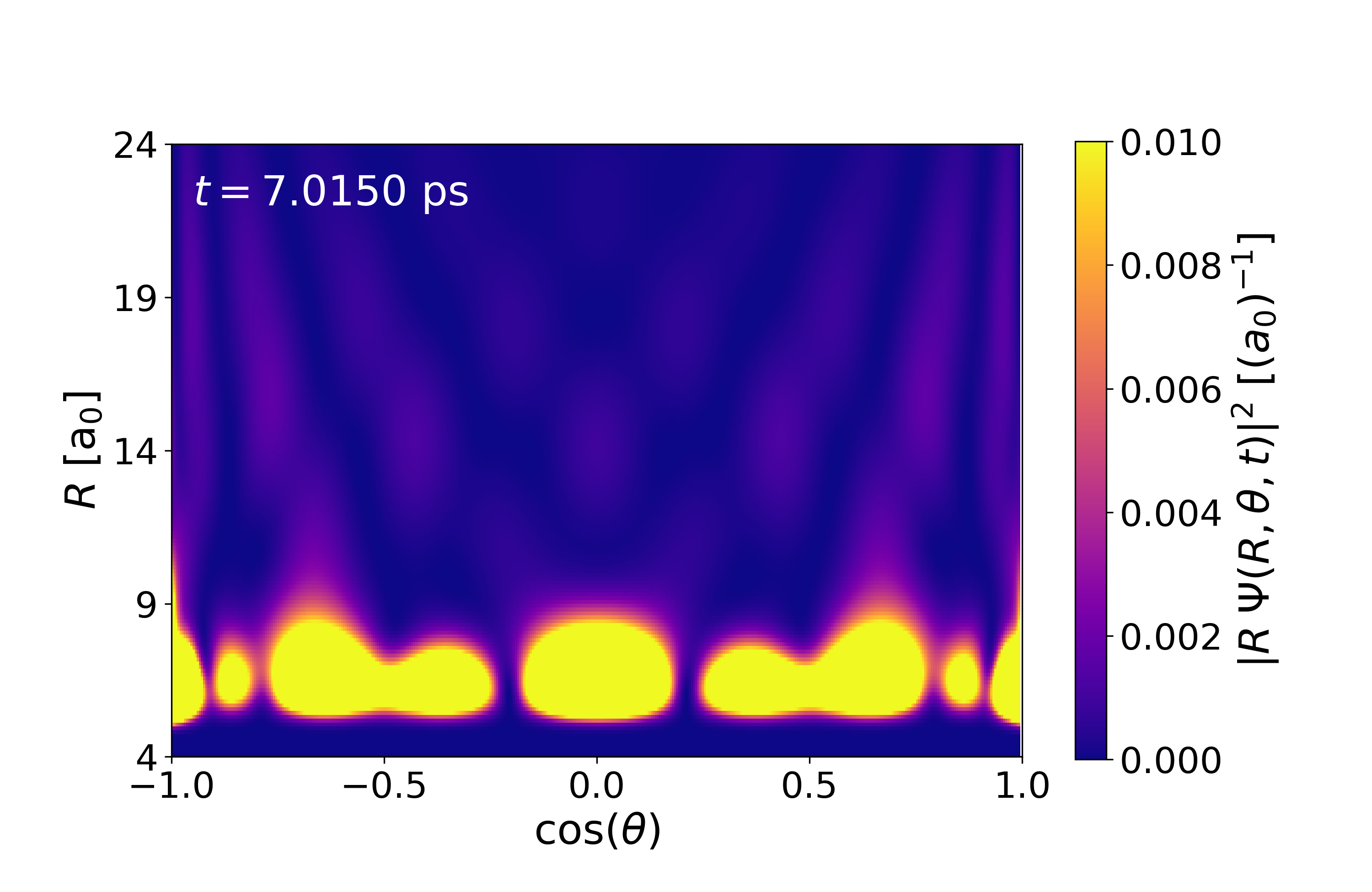}
\vspace*{-0.1in}
    \caption{Illustration of "structured jets."
    The total wave packet density is shown
   for $t=7.0150$~ps and kick strength
    $P_{\text{kick}}=11.9$ as functions of $R$ and $\cos(\theta)$.  
   The maximum of the color scale is set to $0.01$~$a_0^{-1}$ to make the density at larger $R$ visible.
        }
    \label{fig_extra}
\end{figure}

To highlight the intricate structure of the jets, Fig.~\ref{fig_extra} shows a snapshot of the total wave packet density
at $t= 7.0150$~ps. The maximum of the color bar is set to $0.01$~$a_0^{-1}$
(see also the movies in the Supplemental Material~\cite{SM}). On the scale shown, the highly structured nature of the jets can be seen clearly.
Employing fine time and spatial resolutions,  
it should be possible to probe these unbound wave packet portions experimentally.

For $J<14$ [see Fig.~\ref{fig5}(a)-\ref{fig5}(d)],
a portion of the channel density remains at small $R$ out to large times. 
For $t \gtrsim 3$~ps, the $J=0$ density [Fig.~\ref{fig5}(a)] displays oscillatory behavior with a period that 
is set by $E_{\text{vib}}$.  Analogous oscillations can be seen in the $J=2$ and $4$ channels
(for these $J$, the system supports a $v=1$ bound state and the oscillation period is set by the 
energy difference $E_{J,1}-E_{J,0}$). Importantly, the $J=6$ channel density shows similar though
slightly less pronounced oscillatory behavior, despite the fact that the $J=6$ channel does not support a vibrationally excited bound state
(see Table~\ref{tab1}).
For comparison, oscillatory behavior is essentially absent in the $J=8$ channel density for $t \gtrsim 20$~ps.
The oscillatory behavior in the $J=6$ channel is a fingerprint  of the 
resonance state that exists due to the presence of the angular momentum barrier in the effective potential curve $V_{J,\text{eff}}(R)$.
The radial density for the $J=12$ channel does not display any oscillations 
and the density for $R \lesssim 10$~$a_0$ decreases with increasing time. A
fit to an exponential yields $189$~ps. This time scale is 
amenable to experimental verification.

 \section{Summary and outlook}
\label{sec_conclusion}

This paper investigated the distance-dependent dynamics of the neon dimer, induced by a short intense laser pulse. 
The neon dimer is unique among the rare gas dimers, serving as a bridge between the extremely floppy helium dimer and the heavy krypton and xenon dimers.
The neon dimer supports 
$10$ bound states total in the even $J$ channels ($J=0,2,\cdots,10$): 
The low-lying portion of the spectrum is relatively well described 
by the rigid rotor model while the higher-lying portion shows some deviations due to the "opening" of the Born-Oppenheimer potential near the dissociation threshold.
In addition, four of the high-lying states have nodes along the distance coordinate. 
On time scales comparable to the pulse length, a wave packet portion is promoted to unbound "high energy states" and subsequently ejected in 
highly structured "jets." These spatially structured   jets should be observable experimentally.
On longer time scales, the dynamics displays clear fingerprints of the existence of excited resonance states.
One of these, namely the one in the $J=6$ channel, is vibrationally excited.
Vibrationally excited states are completely absent in the rigid rotor model. 
Our work suggests that pump-probe experiments can be used to probe tunneling dynamics that arises due to the 
angular momentum barrier of $V_{J,\text{eff}}(R)$.

Our work motivates experimental work while also suggesting a number of  theoretical follow-up studies. It would be interesting to explore the ro-vibrational dynamics in 
van der Waals dimers with other polarizations, for pulse trains, and for isotope mixtures (e.g., $^{22}$Ne-$^{20}$Ne, where odd-$J$ channels are not disallowed by exchange symmetry). It would also be intriguing 
to investigate the dynamics of van der Waals trimers such as the neon trimer, which may support resonance states that exist thanks to
an angular momentum barrier of effective hyperradial potential curves.

\section{Acknowledgement}
Discussions with Dr. A. Bhowmik are gratefully acknowledged.
D.B. acknowledges support by the National Science Foundation through grant numbers PHY-2110158 and PHY-2409311. Q.G. acknowledges support by the National Science Foundation through grant number PHY-2409600 and support by Washington State University through the Claire May \& William Band Distinguished Professorship Award and the New Faculty Seed Grant. This work used the OU Supercomputing Center for Education and Research (OSCER) at the University of Oklahoma (OU). 
This research was supported in part by grant NSF PHY-1748958 to the Kavli Institute for Theoretical Physics (KITP).
\appendix

\appendix
\section{Rigid rotor model}
\label{sec_appendix_rigid_rotor}
In the rigid-rotor model the internuclear distance $R$ is fixed.
The direction of $\vec{R}$ is parameterized by $\theta$ and $\phi$, where $\theta$ denotes the angle between the molecular axis and the space-fixed $z$-axis and $\phi$ the polar angle in the $xy$-plane.
The
kinetic energy operator $T_{\text{rigid}}$ for a rigid diatomic molecule interacting with a linearly polarized laser 
(polarization vector along the $z$-axis) is
\begin{eqnarray}
T_{\text{rigid}} = 
\frac{ J_{\text{op}}^2}{2 \mu \langle R^2 \rangle} ,
\end{eqnarray}
where $\mu$ denotes the reduced two-body mass ($\mu=m/2$ with $m$ denoting the atom mass)
and $J_{\text{op}}$ the 
angular momentum operator that is associated with the relative
distance vector $\vec{R}$. 
The eigen states of $J_{\text{op}}^2$ are the spherical harmonics $Y_{J,M}(\theta)$ with eigenvalues $\hbar^2 J(J+1)$.
The expectation value $\langle R^2 \rangle$ is taken with respect to the ro-vibrational ground state wave function.

Within the rigid-rotor approximation, the atom-atom potential $V_{\text{aa}}(R)$ reduces to a constant that introduces, under the time evolution, a trivial overall phase. Without loss of generality, we 
set $V_{\text{aa}}(R)=0$.
 The laser-dimer interaction  $H_I(R,\theta)$ reads
\begin{eqnarray}
\label{eq_laser_dimer}
H_I(R,\theta,t)= \nonumber \\
-\frac{1}{2} |\epsilon(t)|^2
\left[ \alpha_{\parallel}(R) \cos^2 \theta + \alpha_{\perp}(R) \sin^2 \theta \right],
\end{eqnarray}
where $\epsilon(t)$ is given in the main text.
Defining 
\begin{eqnarray}
\alpha_{\text{iso}}(R)=\frac{1}{3} \left[ 2 \alpha_{\parallel}(R) + \alpha_{\perp}(R) \right]
\end{eqnarray}  
and
\begin{eqnarray}
\alpha_{\text{aniso}}(R)=\alpha_{\parallel}(R)-\alpha_{\perp}(R)
\end{eqnarray}
and using 
\begin{eqnarray}
Y_{2,0}(\theta)=\sqrt{\frac{5}{16 \pi}} (3 \cos^2 \theta-1),
\end{eqnarray}
Eq.~(\ref{eq_laser_dimer}) 
becomes Eq.~(\ref{eq_laser_molecule}) from the main text.
In the rigid rotor approximation, 
$\alpha_{\text{iso}}(R)$ and $\alpha_{\text{aniso}}(R)$ are replaced by real constants. 
There are different ways of determining the rigid-rotor values of
$\alpha_{\text{iso}}(R)$ and $\alpha_{\text{aniso}}(R)$. Averaging $\alpha_{\text{iso}}(R)$ and $\alpha_{\text{aniso}}(R)$
over the ground state density of the neon dimer yields
\begin{eqnarray}
\overline{\alpha}_{\text{iso}}=\langle \alpha_{\text{iso}} (R)\rangle=-2.544 \times 10^{-3}~a_0
\end{eqnarray}
and
\begin{eqnarray}
\overline{\alpha}_{\text{aniso}}=\langle \alpha_{\text{aniso}} (R)\rangle=0.174 ~a_0.
\end{eqnarray}
The rigid-rotor laser-dimer interaction thus reads
\begin{eqnarray}
\overline{V}_{\text{laser-dimer}}(\theta,t)= \nonumber \\
-\frac{1}{2} |\epsilon(t)|^2
\left[ \overline{\alpha}_{\text{iso}} +  \frac{1}{3} \overline{\alpha}_{\text{aniso}} \sqrt{\frac{16 \pi}{5}} Y_{2,0}(\theta) \right].
\end{eqnarray}

Assuming the initial state is the $J=0$ ground state, the laser only couples states with even $J$ and projection quantum number $M=0$, i.e., the dynamics is independent of $\phi$. 
We expand the time-dependent rigid-rotor wave packet $\Psi(\theta,t)$ as
\begin{eqnarray}
\label{eq_ansatz}
\Psi(\theta,t)= \sum_{J=0,2,\cdots} c_J(t) Y_{J,0}(\theta),
\end{eqnarray} 
where the $c_J(t)$ denote complex expansion coefficients.
Inserting Eq.~(\ref{eq_ansatz}) into the time-dependent
Schr\"odinger equation
\begin{eqnarray}
\imath \hbar \frac{\partial \Psi(\theta,t)}{\partial t}
= \nonumber \\
\left[  \frac{J_{\text{op}}^2}{2 \mu \langle R^2 \rangle} +
\overline{V}_{\text{laser-dimer}}(\theta,t)
\right] \Psi(\theta,t),
\end{eqnarray}
we find a coupled set of first-order differential equations for the
expansion coefficients $c_J(t)$:
\begin{eqnarray}
\imath \hbar \frac{\partial c_J(t)}{\partial t}
= 
\left[ B J(J+1) -\frac{1}{2} |\epsilon(t)|^2 \overline{\alpha}_{\text{iso}} \right] c_J(t) - \nonumber \\
\frac{1}{6} |\epsilon(t)|^2 \overline{\alpha}_{\text{aniso}} \sqrt{\frac{16 \pi}{5}}
\sum_{J'} \langle \langle Y_{J,0}|Y_{2,0} | Y_{J',0} \rangle \rangle c_{J'}(t),
\nonumber \\
\end{eqnarray}
where we defined the rotational constant $B$,
\begin{eqnarray}
B = \frac{\hbar^2} {2 \mu \langle R^2 \rangle}.
\end{eqnarray}
The double brackets denote an angular integral,
\begin{eqnarray}
 \langle \langle Y_{J,0}|Y_{2,0} | Y_{J',0} \rangle \rangle
 = \nonumber \\
 \int_0^{2 \pi} \int_0^{\pi}
 Y_{J,0}^*(\theta) Y_{2,0}(\theta) Y_{J',0}(\theta) 
 \sin \theta d \theta d\phi.
 \end{eqnarray}
It is convenient to define a dimensionless time $\overline{t}$ through
\begin{eqnarray}
\overline{t} = \frac{t}{\hbar/B}.
\end{eqnarray}
With this, the coupled equations become
\begin{eqnarray}
\imath \frac{\partial c_J(\overline{t})}{\partial \overline{t}}
= 
\left[ J(J+1) -\frac{1}{2B} |\epsilon(t)|^2 \overline{\alpha}_{\text{iso}} \right] c_J(\overline{t}) - \nonumber \\
\frac{1}{6B} |\epsilon(t)|^2 \overline{\alpha}_{\text{aniso}} \sqrt{\frac{16 \pi}{5}}
\sum_{J'} \langle \langle Y_{J,0}|Y_{2,0} | Y_{J',0} \rangle \rangle c_{J'}(\overline{t}),
\nonumber \\
\end{eqnarray}
where it is understood that the ratio $t/\tau$ in the expression for $\epsilon(t)$ is evaluated using either the original units for $t$ and $\tau$ or the scaled units for both.

Let us assume that the laser-dimer interaction can be neglected for 
$t \ge \overline{t}_{\text{ref}}$, where $\overline{t}_{\text{ref}}$ denotes some reference time.
At time $\overline{t}_{\text{ref}}$, the expansion coefficients are
$c_J(\overline{t}_{\text{ref}})$. 
Correspondingly, the wave packet for $\overline{t} \ge \overline{t}_{\text{ref}}$
can be written as
\begin{eqnarray}
\Psi(\theta,\overline{t})= \nonumber \\
\sum_{J=0,2,\cdots} c_J(\overline{t}_{\text{ref}}) \exp \left[ -\imath J(J+1) (\overline{t}-\overline{t}_{\text{ref}})
\right] Y_{J,0}(\theta).
\nonumber \\
\end{eqnarray}
The time-dependent expectation value of $\langle \langle \cos^2 \theta \rangle \rangle(\overline{t})$
with respect to the wave packet $\Psi(\theta,\overline{t})$ then becomes
\begin{widetext}
\begin{eqnarray}
\langle \langle \cos^2 \theta \rangle \rangle(\overline{t})
= 
\sum_{J=0,2,\cdots} |c_J(\overline{t}_{\text{ref}})|^2 \langle \langle Y_{J,0}|\cos^2 \theta |Y_{J,0} \rangle
\rangle
+ \nonumber \\
\sum_{J=0,2,\cdots}
2 \langle \langle Y_{J+2,0}|\cos^2 \theta |Y_{J,0} \rangle \rangle \times \\
 \text{Re} \left\{ [c_{J+2}(\overline{t}_{\text{ref}})]^*  c_J(\overline{t}_{\text{ref}}) 
\exp \left[ \imath ((J+2)(J+3)-J(J+1)) (\overline{t}-\overline{t}_{\text{ref}}) \right] \right\},
\nonumber \\
\end{eqnarray}
\end{widetext}
where we used that only matrix elements with $|J'-J|=0$ and $2$ give non-zero contributions.
The quantity $(J+2)(J+3)-J(J+1)=2(2J+3)$ takes the values
$6$, $14$, $22$, and $30$ for $J=0$, $2$, $4$, and $6$, respectively.
If we assume that several $J$ states are occupied at $\overline{t}=\overline{t}_{\text{ref}}$, then we
see that we get identical expectation values at $\overline{t}_{\text{ref}}$ and $\overline{t}_{\text{ref}}+ \pi$, i.e., we expect a full revival after the dimensionless time interval $\overline{T}=\pi$ or, in dimensionful units, after the time interval $T=\pi \hbar /B$.

For the neon dimer, we have
 $\sqrt{\langle R^2 \rangle}=6.320$~a$_0$
 (for comparison,
the
expectation value of the internuclear distance $R$ in the ro-vibrational ground state is less than $0.5$~\% smaller, namely
$\langle R \rangle =6.293$~a$_0$) and 
\begin{eqnarray}
B^{-1}= \frac{2 \mu \langle R^2 \rangle}{\hbar^2} = 1.456 \times10^6~\text{a.u.},
\end{eqnarray}
which corresponds to
\begin{eqnarray}
B=6.870 \times 10^{-7}~\text{a.u.}=0.2170~\mbox{K}.
\end{eqnarray}
It follows
\begin{eqnarray}
T=\frac{\pi \hbar}{B}=4.573 \times 10^6~\text{a.u.}=110.6~\mbox{ps}.
\end{eqnarray}


\begin{thebibliography}{98}
\bibitem{footnoteADK}
The ionization probability for $\tau=933$~fs and $I=1.17 \times 10^{14}$~W/cm$^2$ is $\approx 0.1$~\% for 
linearly polarized laser light with $\lambda=800$~nm. This estimate is obtained using the ADK
formalism~\cite{ADK}.
For fixed $\tau$, the ADK rate decreases rapidly with decreasing $I$.
\bibitem{footnote1}
Shortening the pulse length while keeping the kick strength constant leads to an enhanced ionization probability. Since the ionization channels are excluded from our formulation, the numerical treatment allows for a shortening of the pulse length. In experiment, the limit of the shortest $\tau$ is set by the requirement to keep the ionization probability negligibly small.
See also the comments in Ref.~\protect\cite{ footnoteADK}.
\end{thebibliography}
\end{document}